# Radiation dose estimation in pencil beam x-ray luminescence computed tomography imaging

Ignacio O. Romero, and Changqing Li*
Department of Bioengineering, University of California, Merced, Merced, CA, USA.

*Corresponding Author: Changqing Li, Tel.: (209) 228-4777; Email: cli32@ucmerced.edu

**Abstract:**

Pencil x-ray beam imaging provides superior spatial resolution than other imaging geometries like sheet beam and cone beam geometries due to the illumination of a line instead of an area or volume. However, the pencil beam geometry suffers from long scan times and concerns over dose discourage laboratory use of pencil beam x-ray sources. Molecular imaging techniques like XLCT imaging benefit most from pencil beam imaging to accurately localize the distribution of contrast agents embedded in a small animal object. To investigate the dose deposited by pencil beam x-ray imaging in XLCT, dose estimations from one angular projection scan by three different x-ray source energies were performed on a small animal object composed of water, bone, and blood with a Monte Carlo simulation platform, GATE (Geant4 Application for Tomographic Emission). Our results indicate that, with an adequate x-ray benchtop source with high brilliance and quasi-monochromatic properties like the Sigray source, the dose concerns can be reduced. With the Sigray source, the bone marrow was estimated to have a radiation dose of 30 mGy for a typical XLCT imaging, in which we have 6 angular projections, 100 micrometer scan step size, and $10^6$ x-ray photons per linear scan.

## 1. Introduction

Pencil x-ray beam imaging provides superior spatial resolution than other x-ray imaging geometries like sheet beam and cone beam geometries because the fine pencil beam can provide anatomical guidance in the image reconstruction. However, a pencil beam scan suffers from long scan times and concerns over dose discourage laboratory use of pencil beam x-ray sources. Molecular imaging techniques like narrow beam x-ray luminescence computed tomography (XLCT) imaging benefit most from pencil beam imaging. XLCT combines high measurement sensitivity of optical imaging and high spatial resolution of x-ray imaging. XLCT uses x-ray beams to excite nanophosphors, which emit optical photons to be measured for optical imaging. The width and position of the pencil beam is used as structural guidance to reconstruct the XLCT image therefore creating a molecular image with high spatial resolution [1-3]. The structural guidance information is lost when using the sheet beam and cone beam geometries for XLCT imaging which limit the sensitivity and spatial resolution [4, 5-8]. To minimize the dose, benchtop x-ray sources are filtered to remove the lower energy x-ray photons, but the added filtration increases the scan times.

In this work, the dose deposited into a small object (about 5 mm in diameter) using pencil x-ray beam geometries was simulated in GATE (Geant4 Application for Tomographic Emission) [9]. The small object is composed of water, bone and blood to simulate the bone marrow structures of a mice leg. Three different x-ray sources were simulated, and the dose and dose rate delivered from one angular projection scan is shown and compared. The dose measurements were repeated with different factors of the x-ray output rates to generate the linear relationship of the absorbed dose in each structure to the x-ray source output photon number. Among the three different x-ray sources, a benchtop quasi-monochromatic x-ray source from Sigray Inc was modeled.

The paper is organized as follows. In section 2, the methods of the GATE simulation, setup, and source spectra are presented. In section 3, the results showing the relationships between x-rays output and dose with different spectra are presented. The paper concludes with a discussion of the results and future works.

## 2. Methods

### 2.1 GATE programming and setup

The GATE software is a GEANT4 wrapper which utilizes the macro language to ease the learning curve of GEANT4 and allow GEANT4 to be more accessible to researchers [9]. The GATE simulations in this work were parallelized and executed with a custom bash script on a 20 CPU workstation. The Penelope physics package (empenelope) from Geant4 was used to model the physical processes. The Dose Actor feature from GATE was used to store the dose delivered to the volume inside a 3D matrix [10].

A 5 mm diameter cylindrical water phantom with a 2 mm diameter cylindrical bone structure located at the phantom center was simulated in GATE. A 1 mm diameter cylindrical blood structure was centered in the bone structure to simulate the bone marrow. The imaging object was placed at the center of the reference axis. The source beam width at the central coronal slice (x-axis) is 100 $\mu$m. A linear scan step size of 100 $\mu$m was used to acquire a single angular projection. The schematic of the GATE simulation setup along with a snapshot of the GATE simulation is shown in Fig. 1. The black box in the schematic is a representation of the dose actor tool from GATE which is seen as the white framed box around the simulated object.

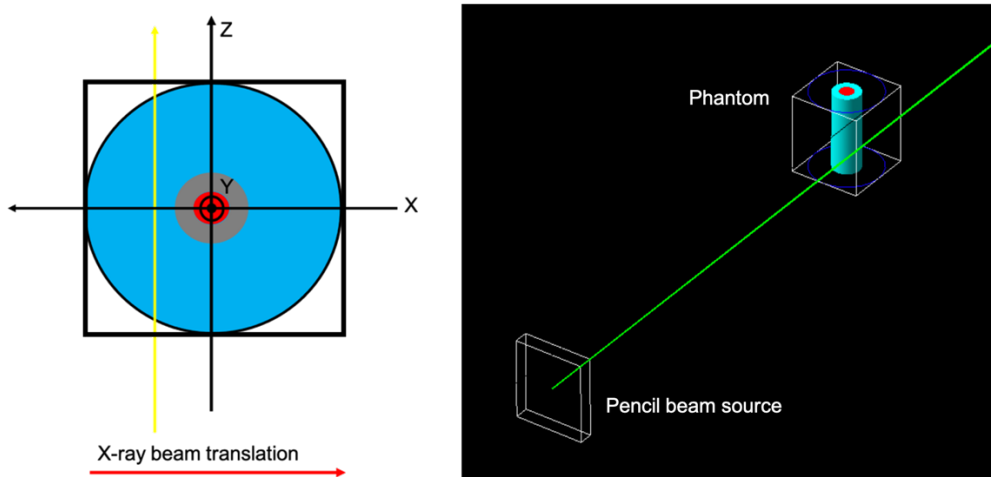

Figure 1: Schematic (left) and simulation snapshot (right) of the pencil beam dose GATE simulations.

The dose was stored in a 5 mm$^3$ cube with voxel size of 0.1 mm$^3$. 5×10$^5$, 10$^6$, and 1.5×10$^6$ x-ray photons were used for each linear scan step to establish a linear relationship between x-ray photon number and dose. From the linear relationship, the dose delivered by the x-ray sources at any output rate may be estimated by linear interpolation.

*2.2 Source Spectra Modeling in GATE*

To simulate the Sigray source, its energy spectrum was estimated and plotted in Fig. 2. To simulate the XOS x-ray source (X-Beam Powerflux [Mo anode], XOS) in our lab, a 50 kVp spectrum of the x-ray source with polycapillary lens was acquired and normalized. The lab x-ray tube focuses x-rays to an approximate focal spot size of 100 μm at its focal distance. The histogram user spectrum tool from GATE was used to import the XOS source histogram. The energy of the emitted photon within each bin of the histogram was distributed uniformly. The spectrum is plotted in Fig. 3a below. Due to the utilization of a polycapillary lens, the spectrum was truncated after 30 keV. To minimize the dose contribution from low energy x-ray photons, a 2 mm Al filter was positioned in front of the source. The spectrum for the filtered XOS lab source is calculated and plotted in Fig. 3b.

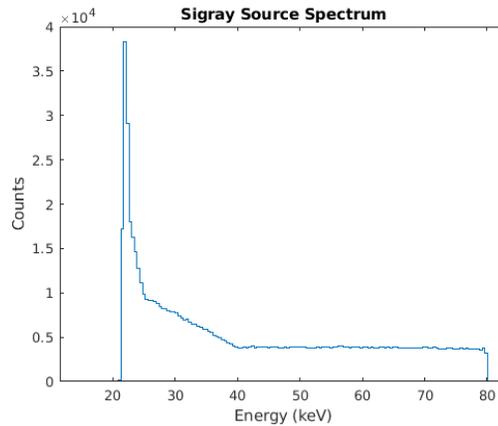

Figure 2: The x-ray energy spectrum of the x-ray source from Sigray, Inc.

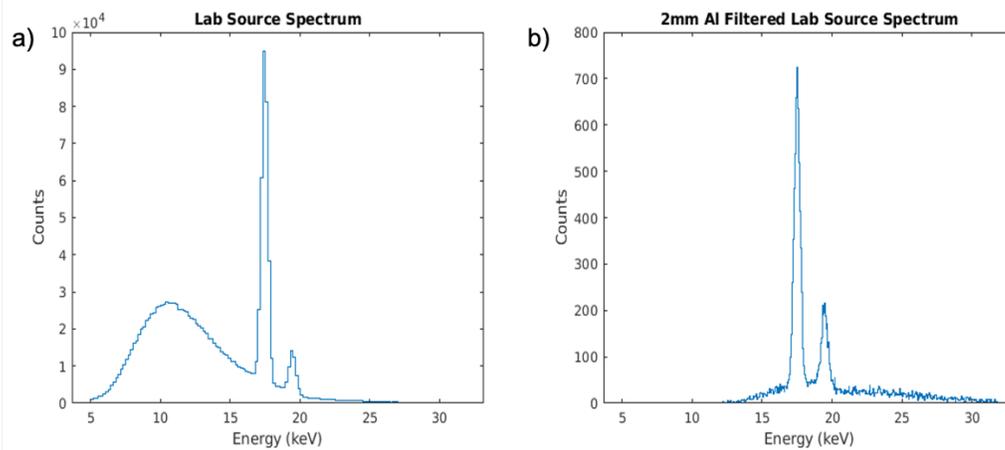

Figure 3: Modeled XOS lab source spectrum without (a) and with (b) a 2 mm thick aluminum filter after the polycapillary lens.

*2.3 Dose estimations*

The dose stored in each voxel in GATE is calculated as the energy deposited per mass of the voxel. The dose for bone marrow, bone, and object background were calculated as the mean dose from each region. This was done by using masks to identify only those pixels belonging to the bone marrow, bone, or object background and then the values in each region were averaged. All dose calculation was done in MATLAB after GATE simulations.

## 3. Results

### 3.1 Results of dose map for one projection XLCT imaging

Fig. 4 shows the central line profile and dose maps for the cases of the filtered XOS source, unfiltered XOS source, and Sigray source. All the sources had an x-ray photon number of $10^6$ per linear scan step. All three cases had 50 linear steps with the same pencil beam and step size of 100 micrometers. From Fig. 4, it is shown that the bone structure absorbed more dose than the surrounding background and bone marrow structures due to the bone's greater effective atomic number (Z). In Fig. 2a, the addition of the Al filter results in the removal of the high entrance dose seen in Figure 2b. Without the Al filter, the lower energies from bremsstrahlung result in increased air ionizations and superficial dose which result in the high entrance dose as seen in the central slice line profile and dose map of Fig. 2b. However, by adding a filter, the mean energy of the x-ray source is increased which results in beam hardening and the x-rays become more penetrating which deliver greater dose to deeper structures. This can be seen on by the increased dose in the right bone peak and the exit dose of the central slice line profiles of Fig. 2a compared to Fig. 2b. The shadowing effects are also less severe in Fig. 2a due to the beam hardening effects. Fig. 2c shows more uniformity in the dose distribution due to a greater average energy by the Sigray source and less dose in the bone morrow region compared with the XOS source cases.

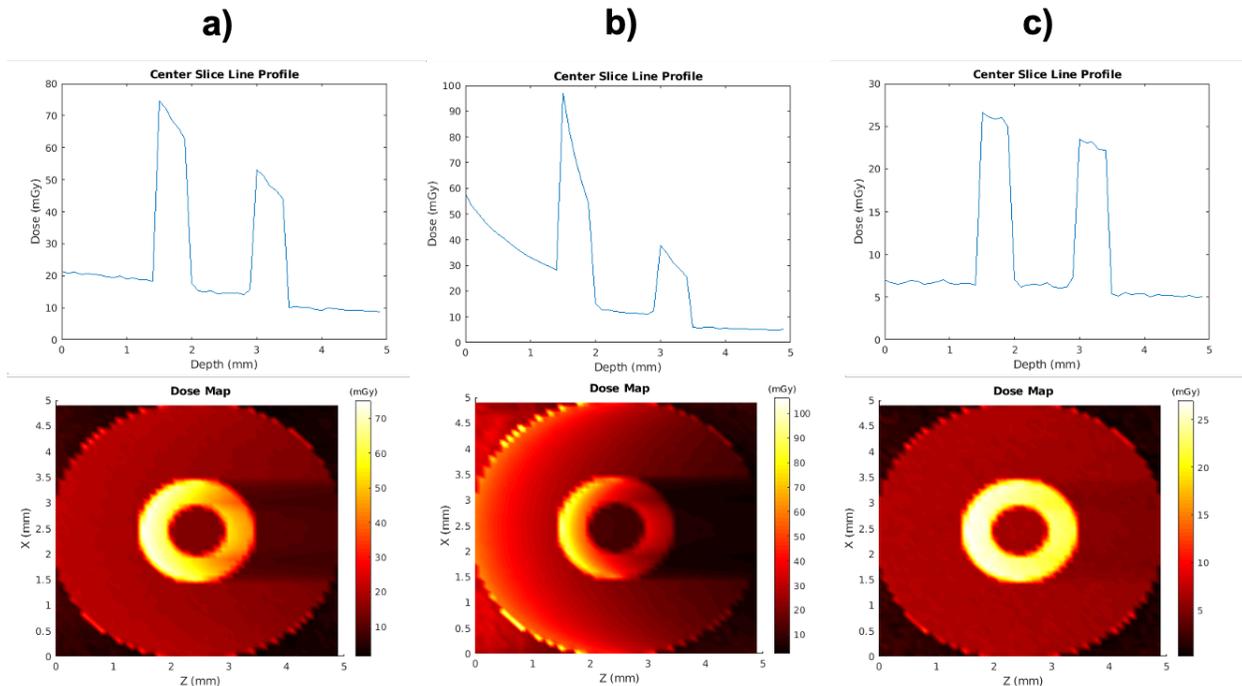

Figure 4: Center slice line profile and dose maps of (a) Filtered XOS source, (b) Unfiltered XOS source, (c) Sigray source for $10^6$ x-ray photons per linear scan step for all 50 steps.

*3.2 Dose estimation based on x-ray photon number for one projection XLCT imaging*

Fig. 5 shows the regression line plots with trendline equations relating the x-ray number per linear scan step (photons/step) to the absorbed average dose (Avg Dose) in milliGray (mGy) units for cases of the filtered XOS source, the unfiltered XOS source, and the Sigray source and for the different components of the phantom. In Fig. 5a, three points were plotted for the bone marrow and bone structure to show the high linearity between the x-ray photon number and absorbed dose as shown by the $R^2$ value of 1. For Figs. 5b and 5c, only two points were plotted using $5\times10^5$ and $10^6$ x-ray photons. For all regression plots, the bone structure shows greater values of average absorbed dose than the bone marrow and background structures. All the trendlines equations for the bone structure show a greater slope therefore bone will absorb dose at a greater rate than the other structures. For Fig. 5a, the regression lines of bone marrow and background are identical. Fig. 5a shows a slight increase in the dose rate of the bone structure due to the beam hardening effects from the Al filter. In Fig. 5b, the background regression curve is greater than the bone marrow regression curve due to the bremsstrahlung energies depositing significant dose to the background. Fig. 5c shows the lowest absorbed dose rate for the bone structure. Fig. 5c also shows a decrease of at least a half order of magnitude in the absorbed dose rate in bone marrow and background structures.

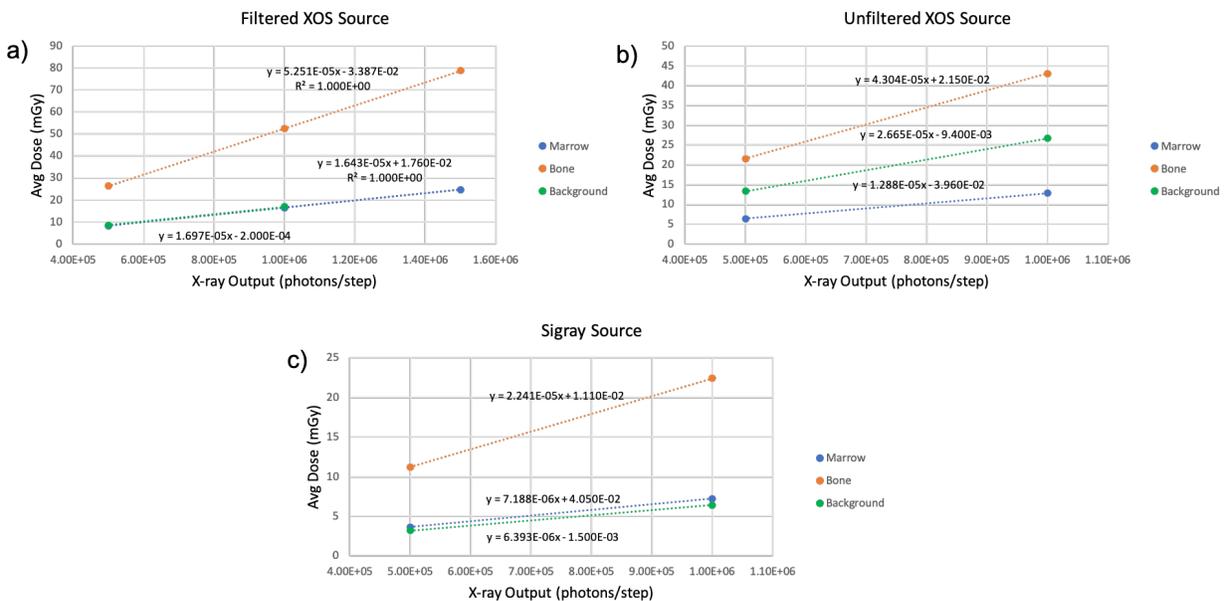

Figure 5: Regression line plots of (a) filtered XOS source, (b) unfiltered XOS source, and (c) Sigray source for the different components of the phantom. In (a), the marrow and background plots are overlaid.

*3.3. Dose estimation of a typical XLCT scan*

As we reported in our XLCT imaging studies [11-13], we only need 6 angular projections for the narrow beam based XLCT imaging. The x-ray tube output photon number in each linear scan step depends on many factors including the detector sensitivity, the x-ray excitable phosphor particle brightness, the linear scan speed, and the target size and depth, etc. Generally, we can assume the x-ray photon number to be about $10^6$ per linear scan with a scan step size of 100 micrometers. For each angular projection, the radiation dose in bone marrow was calculated to be 6 mGy as indicated in Fig. 5c for the Sigray source case. For a typical XLCT scan with 6 angular projections, we estimate the bone marrow dose to be about 30 mGy.

## 4. Discussions and Conclusions

In this work, the dose deposited into a 5 mm diameter object by a pencil x-ray beam geometry was simulated in GATE. The dose deposited by three different x-ray source spectra was shown and compared. The removal of bremsstrahlung energies results in decreased superficial dose thus making the filtered XOS source or the Sigray source a better choice for imaging.

Higher mean x-ray energies were observed to give greater dose in deeper depths due to beam hardening effects. This effect was observed when comparing the unfiltered XOS source to the filtered XOS source results. With the filtered XOS source, the bone structures and bone marrow had greater dose rate as seen in the regression plot in Fig. 5. X-rays becomes less susceptible to the photoelectric effect at greater energies so the x-rays may be fully absorbed at greater depths which explains the greater background dose and bone dose with the filtered XOS source.

Due to the size of the imaging object, the Compton scattering dose effects are minimal compared to the photoelectric effect. However, the photoelectric effect is inversely related to the energy of the x-ray. Therefore, at much greater energies, x-ray transmission is more likely to occur. This explains why the filtered XOS source shows greater dose and dose rates for all structures than the Sigray source. The Sigray source would be a great alternative to standard lab sources in XLCT imaging. With the Sigray source, more transmitted x-rays will be absorbed to excite the phosphor contrast agents while the surrounding tissues are spared from dose.

In summary, dose measurements were performed on a 5 mm diameter object size with three different pencil beam x-ray spectra. This work provides insight to the concerns associated with XLCT imaging using a pencil beam geometry. With an adequate x-ray benchtop source with high brilliance and quasi-monochromatic properties like the Sigray source, the dose concerns can be reduced. With the Sigray source, the bone marrow was estimated to have a radiation dose of 30 mGy for a typical XLCT imaging with 6 angular projections and 100 micrometer scan step size and $10^6$ x-ray photons per linear scan. The findings of this work will also be beneficial to other molecular x-ray imaging modalities like x-ray fluorescence computed tomography if it relies on a pencil x-ray beam imaging geometry. In future works, a small mouse model will be imported to GATE to estimate the dose in a XLCT imaging of small animals.

## Acknowledgements

This work was funded by the NIH National Institute of Biomedical Imaging and Bioengineering (NIBIB) [R01EB026646]. Authors thanks Sigray, Inc. for providing the spectral information of the Sigray x-ray source.## 5. References

[1] Li, C., Martínez-Dávalos, A., & Cherry, S. R. (2014). Numerical simulation of x-ray luminescence optical tomography for small-animal imaging. *Journal of Biomedical Optics*, *19*(4), 046002.